# $T_e/T_i$ effects on JET energy confinement properties

E Asp[†‡], J Weiland[§], X Garbet[†], P Mantica[∥], V Parail[¶],
W Suttrop[+] and the EFDA-JET contributors

[†] Assoc. Euratom-CEA sur la fusion, CEA Cadarache, 13108 St Paul-Lez-Durance, France
[§] Chalmers University of Technology, EURATOM-VR Fusion Assoc., 412 96 Göteborg, Sweden
[∥] Istituto di Fisica del Plasma, EURATOM-ENEA-CNR, via Cozzi 53, 20125 Milan, Italy
[¶] EURATOM/UKAEA, Culham Science Centre, Abingdon, OX14 3DB, United Kingdom
[+] MPI für Plasmaphysik, IPP-EURATOM Assoz., 8046 Garching bei München, Germany

**Abstract.**
Lately the question has been raised if a modification of the energy-confinement scaling law with respect to the electron to ion temperature ratio, $T_e/T_i$, is required. Theoretically, like in *e.g.* the Weiland model, the confinement is thought to degrade with $T_e/T_i$ and studies of the hot-ion ($T_i > T_e$) mode seems to corroborate this. In this paper, it is shown that due to a number of effects that cancel each other out, the energy confinement time remains constant for $T_e/T_i \gtrsim 1$. The numerical study relies on a series of JET shots specifically designed to reveal an effect of $T_e/T_i$ in the hot-electron ($T_e > T_i$) mode. A distinct effort was made to keep all current scaling-law parameters constant, including the total heating power. The effects that provide the constant confinement times have therefore nothing to do with the global properties of the plasma, but are rather due to variations in the temperature gradients which affects the transport locally.

## 1. Introduction

The common JET scenario of today contains a significant amount of neutral beam injection (NBI) heating to improve the plasma energy confinement time by entering the hot-ion mode[1][2]. In a burning plasma there is also an other advantage of a high ion temperature as this boosts the fusion reaction rate. These beneficial effects of the hot-ion mode can be lost in an ignited plasma where the $\alpha$-particles created in the fusion reaction mainly heats the electrons. It is hence important to explore what might happen to the energy confinement in the event of an $\alpha$-particle induced hot-electron plasma. Thermal equilibration of the temperatures in an ITER-like plasma with high density will quite likely be very efficient and will bring the electron to ion temperature ratio, $T_e/T_i \sim 1$[3].

Current scaling laws[3] used for plasma energy confinement do not contain any dependence on the temperature ratio or applied power ratio. This is quite remarkable since, as mentioned above, a clear improvement of confinement has been found for low $T_e/T_i$. Since the confinement in JET for high values $T_e/T_i$ is less investigated, this work questions the validity of the current scaling laws in the hot-electron mode. Suttrop *et al* devoted a series of JET discharges to this end[4]. These discharges are very similar from a scaling-law perspective and, in principle, it ought to be a straightforward task to extract the temperature ratio dependence. It should be noted that even the total applied power was constant and that the obtained ratios of $T_e/T_i$ were accomplished by varying the ratio of electron to ion input power. Quite surprisingly, the energy confinement times appeared to be impervious to any change in $T_e/T_i$.

[‡] E-mail: Elina.Asp@cea.fr

Superficially, the temperature dependencies of the Weiland model[5][6][7] seems to predict better confinement for small $T_e/T_i$ and that the confinement degrades as the temperature ratio increases. This derives from the stabilization of the ion temperature gradient (ITG) mode at low $T_e/T_i$ and from that the trapped electron (TE) mode amplifies at higher $T_e$. Now, what ultimately determines the confinement time is the effective diffusivity derived from the heat fluxes. Hence, it is the weighted average of the each diffusivity with respect to their species temperature gradient that yields the final confinement time. No variation of the latter would be observed if the changes in the gradients counteract the growth or decline of their respective diffusivity. As it turns out, this is exactly what happens in the case of the ion instability. For the electrons, it is a combination of the lack of $T_i$ dependence and fluctuating boundary conditions that results in a flat electron contribution to the total effective diffusivity. With the electron and ion transports above leading to only a weak growth of the total effective diffusivity, no trend in the energy confinement time can be seen. It has also been shown that the H-factors of the studied shots are not influenced by either $T_e/T_i$ or the absorbed electron to ion power ratio[8]. Thus, no correction of the current scaling laws is needed for $T_e/T_i > 1$.

This paper begins with the crucial theory for understanding the lack of $T_e/T_i$ dependence in section 2, followed by the experimental and numerical results obtained by the JETTO code[9] in section 3. Finally, conclusion are drawn in section 4.

## 2. Theoretical Overview

Current scaling laws do not include any dependence on temperature or applied power ratio, as for *e.g.* the empirical scaling law best suited to describe the kind of shots presented in this paper[3],

$$\tau_{\text{IPBA98(y,2)}} = 0.0562 I_\text{P}^{0.90} B_0^{0.15} P^{0.69} M^{0.19} R^{1.97} n^{0.41} \epsilon^{0.58} \kappa_\text{a}^{0.78}. \quad (1)$$

The notations used above are, $I_\text{P}$-plasma current, $B_0$-toroidal magnetic field,
$P = P_\text{abs} - \partial W_\text{th}/\partial t$, $P_\text{abs}$-absorbed heating power, $W_\text{th}$-thermal energy content,
$M$-mass number of the fusion ions, $R$-major radius, $\epsilon = a/R$, $a$-minor radius and
$\kappa_\text{a} = (\text{poloidal cross-section})/\pi a^2$. This is quite remarkable considering that theoretical models describing the transport in the bulk plasma have important temperature dependencies.

In the Weiland model the diffusivities driven by either the ITG or the TE mode yield for the ions

$$\chi_i = \frac{\gamma_i}{k_r^2} \propto T_e T_i^{1/2} \sqrt{\frac{R}{L_{T_i}} - \frac{R}{L_{T_i,\text{th}}}} \qquad (2)$$

and the electrons

$$\chi_e = \frac{\gamma_e}{k_r^2} \propto T_e^{3/2} \sqrt{\frac{R}{L_{T_e}} - \frac{R}{L_{T_e,\text{th}}}}, \qquad (3)$$

if we assume that the modes are uncoupled and that the radial correlation length is of the same order as the poloidal one, *i.e.* $k_r^2 \approx k_\theta^2 = 0.1/\rho_s^2$. This derives from $k_\theta^2 \rho_s^2 = 0.1$ ($\rho_s$ is the ion Larmor radius at the electron temperature) for the most unstable modes[10]. The JETTO code uses this constant value and hence it is regarded as fixed in the analytical expressions too. In equations (2) and (3) the subscripts i=ion and e=electron, $\gamma$ is the growth rate, $T$ is the temperature, $R$ is the major radius, $L_T$ is the temperature inhomogeneity length scale and the thresholds are given by

$$\frac{R}{L_{T_i,\text{th}}} = \frac{2}{3}\frac{R}{L_n} + \frac{10}{9}\frac{T_i}{T_e} \qquad (4)$$

and

$$\frac{R}{L_{T_e,\text{th}}} \neq f(T_e, T_i), \qquad (5)$$

where $L_n$ is the density inhomogeneity scale length. In this basic model valid for $R/L_n < 2$, the TE mode lacks any dependence on ion parameters and temperature. The ion diffusivity (2) has two features which enhance transport when $T_e/T_i$ increases. First of all the magnitude is proportional to $T_e$, secondly the ion threshold (4) is inversely proportional to $T_e/T_i$ and becomes lower. A higher electron to ion temperature ratio is either accomplished by increasing the electron heating or decreasing the ion heating. Since both cases lead to more ion heat transport without balancing it with more applied ion power, the ion temperature drops and its profile flattens. This in turn leads to even higher values of $T_e/T_i$ which again augments $\chi_i$. It is the ion threshold the provides the feedback to refrain this amplification mechanim to blow up, by setting a lower limit to $R/L_{T_i}$. Below the threshold the ion transport is effectively cut off. A drop in the ion temperature like the one described here, has been observed in both DIII-D[11] and AUG[12] for increasing electron heating.

One of the answers to why no dependence on $T_e/T_i$ has been added to the energy confinement time (1) might be this process for the ions, which simultaneously augments the diffusivity and flattens the temperature profile. The confinement depends on both of these parameters through the total effective diffusivity,

$$\chi_{\text{tot}}^{\text{eff}} = \frac{\chi_e \partial_r T_e + \chi_i \partial_r T_i}{\partial_r T_e + \partial_r T_i}, \qquad (6)$$

where $\partial_r = \partial/\partial_r$. The ion contribution to the above equation may simply not change much as $T_e/T_i$ rises. On the other hand, the electron part ought to increase $\chi_{\text{tot}}^{\text{eff}}$ as the electron transport and temperature gradient are expected to increase for larger amounts of electron heating. Hence the question remains, is there any justification for not adding a $T_e/T_i$ dependence in the scaling laws for $T_e/T_i \gtrsim 1$.

## 3. Results

The results presented in this section were obtained by numerical simulations using the Weiland model as provided by the JETTO code. The $T_e/T_i$ scaling of the plasma energy confinement time are shown and explained by studying the bulk heat diffusivities. To strengthen the correlation between global and local parameters, the diffusivities and temperatures were evaluated at the normalized minor radius $\rho = 0.6$, which lies well outside of the heating region. Figure 1 shows no trend of the confinement times *vs.* $T_e/T_i$ in the

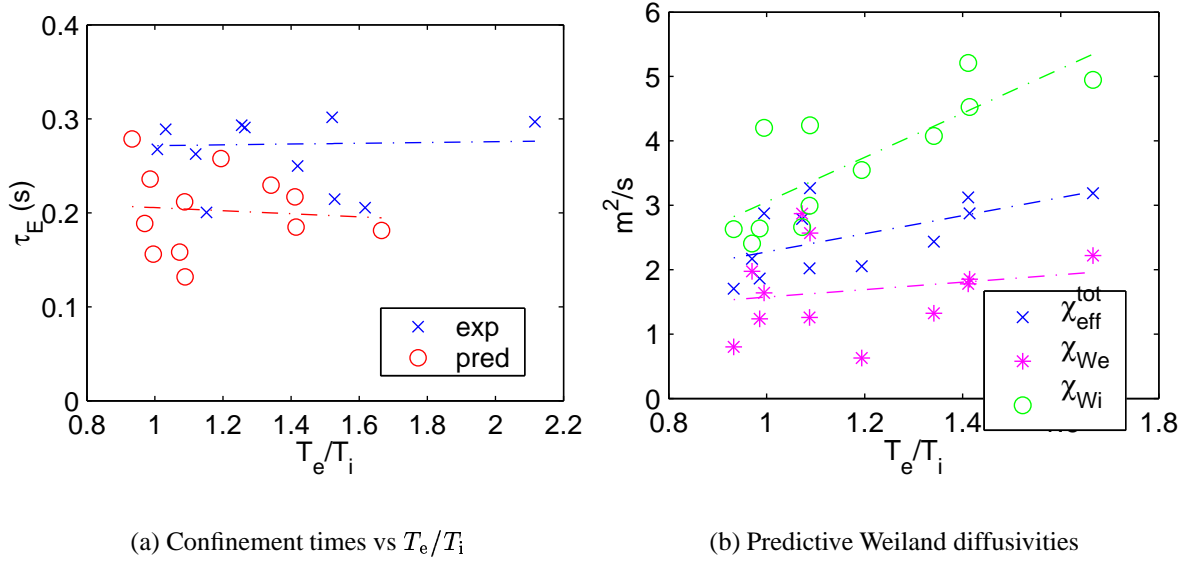

(a) Confinement times vs $T_e/T_i$   (b) Predictive Weiland diffusivities

**Figure 1.** The weak trend of $\chi_{\text{eff}}^{\text{tot}}$ is reflected in the lack of $T_e/T_i$ scaling of $\tau_E$.

studied interval. This holds both for the experimentally (exp) and numerically predicted (pred) values and is in accordance with the empirical scaling law (1). The total effective diffusivity only displays a weak trend with $T_e/T_i$, and it is moreover quite scattered. Despite the magnitude of the electron diffusivity being smaller than the ion diffusivity, it governs the overall behavior of the effective diffusivity. When $T_e/T_i$ increases the ion temperature gradient decreases due to the process discussed in section 2 and keeps the ion contribution to the total diffusivity (6) almost constant. The growing, although scattered $\partial_r T_e$ adds to the importance of $\chi_e$. That $\chi_e$ shows a weak trend with $T_e/T_i$ is expected. According to equations (3) and (5) the pure electron branch does not contain any $T_i$ dependence and $T_e$ does not necessarily rise to obtain higher values of $T_e/T_i$. Making a gyro-Bohm normalization and plotting $\chi_e/T_e^{3/2}$ *vs.* $T_e$ does not decrease the scattering of the data (figure 2(a)), as expected if the electron diffusivities (3) were mostly driven by the amplitude $\sim T_e^{3/2}$. The variation in the electron diffusivity arises rather from the height-above-threshold (figure 2(b)). As the density profiles are fairly flat for all shots except the one which strongly deviates from the rest in figure 2(a), the thresholds given by (5) do not vary much either. Therefore, it must be the values of $R/L_{T_e}$ that drives the electron transport and induces the scattering. These

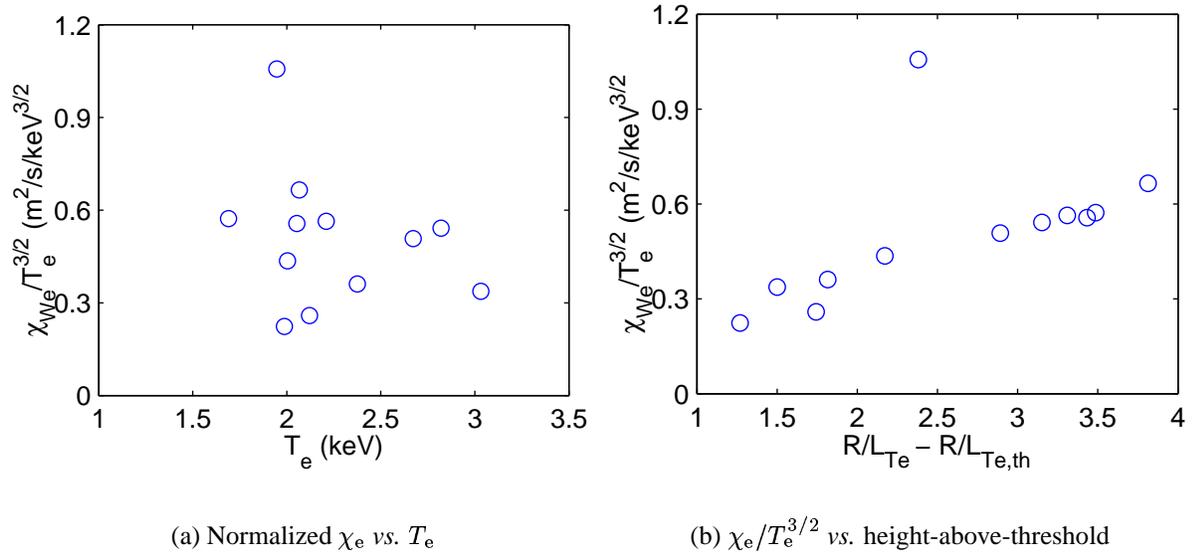

(a) Normalized $\chi_e$ vs. $T_e$

(b) $\chi_e/T_e^{3/2}$ vs. height-above-threshold

**Figure 2.** The reason for the scattering of $\chi_e$ can be found in the height-above-threshold. Only in the case of a coupled TE mode ($R/L_{ne} \approx 4$) the deviation from this trend is significant

values are strongly correlated to the boundary electron temperatures (figure 3(a)). The spread of the boundary $T_e$ vs. local $T_e/T_i$ in figure 3(b) is substantial and hence we conclude that the scattering of the confinement times in figure 1(a) derives from the variation in the boundary temperatures of the electrons.

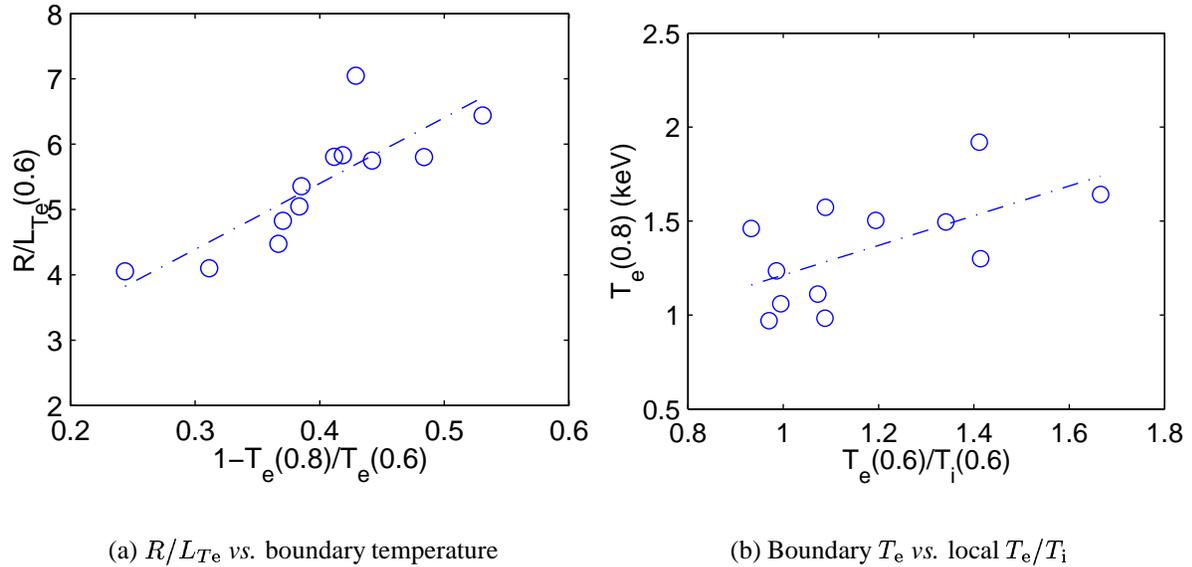

(a) $R/L_{Te}$ vs. boundary temperature

(b) Boundary $T_e$ vs. local $T_e/T_i$

**Figure 3.** The boundary temperature has been chosen at the beginning of the validity region of the Weiland model, *i.e.* at $\rho = 0.8$. A simple estimate, $\partial_r T_e(0.6) = (T_e(0.6) - T_e(0.8))/0.2$ was used to retrieve the dependence on $T_e(0.8)$. The change in $T_e(0.8)$ affects $\partial_r T_e$ in a way that the same $T_e$ (or $T_e/T_i$) yields a range of $R/L_{Te}$ which creates a spread in $\chi_e$.

## 4. Conclusions

The results of this paper yield no motive to insert a temperature ratio dependence in the current plasma energy confinement scaling laws for $T_e/T_i \gtrsim 1$. Although both $T_e$ and $T_i$ are important for the transport on a microscopic level, their impact on the trend of the global confinement in this case becomes negligible. The reason for this is two-fold. Firstly, the ion temperature profiles flatten and drop as $T_e/T_i$ increases, which make the ion contribution to the total effective diffusivity fairly constant. Secondly, the boundary temperatures of the electrons fluctuates in-between the shots, giving a range of $\nabla T_e$ for similar values of $T_e$. As the thresholds for the onset of the trapped electron instability are more or less the same for all the shots, diverse amounts of transport are obtained at approximately equal $T_e/T_i$. This translates into a large scatter of the confinement times. In future experiments the second point could be easily circumvented by having equivalent edge temperatures for a series of shots. The first point however, is more difficult to suppress. The drop in the ion temperature that follows an augmentation of $T_e/T_i$ is an intrinsic property of the ion temperature gradient mode with $R/L_n \sim 2$. In the limit of very flat density profiles, it might be prevented since the threshold then becomes proportional to $T_e/2T_i$ in addition to the $20T_i/9T_e$ presented here.